\documentstyle[12pt,amsmath,amsfonts, graphicx]{article}
\newcommand{\lC}{\mathrm{l\hspace{-2.1mm}C}}
\newcommand{\lR}{\mathrm{I\hspace{-0.7mm}R}}

\numberwithin{equation}{section}

\begin{document}

\hoffset = -1truecm
\voffset = -2truecm

\begin{titlepage}
{\flushright
\today
\\}
\vskip 1truecm {\center\LARGE \bf Deformation of Curved BPS Domain
Walls and Supersymmetric Flows on $2d$ K\"ahler-Ricci Soliton
\\}

\vskip 2truecm

{\center \bf
 Bobby E. Gunara and Freddy P. Zen \footnote{email: bobby@fi.itb.ac.id,
 fpzen@fi.itb.ac.id}
\\}

\vskip 1truecm

\begin{center}
\textit{Indonesia Center for Theoretical and Mathematical Physics
(ICTMP),
\\
and \\
Theoretical  Physics Laboratory, \\
Theoretical High Energy Physics and Instrumentation Division,\\
Faculty of Mathematics and Natural Sciences, \\
Institut Teknologi Bandung\\
Jl. Ganesha 10 Bandung 40132, Indonesia.}

\vskip 0.5truecm

\end{center}

\vskip 1truecm

{\center \large \bf
ABSTRACT
\\}
\vskip 1truecm

\noindent We consider some aspects of the curved BPS domain walls
and their supersymmetric Lorentz invariant vacua of the four
dimensional $N=1$ supergravity coupled to a chiral multiplet. In
particular, the scalar manifold can be viewed as a two dimensional
K\"ahler-Ricci soliton generating a one-parameter family of
K\"ahler manifolds evolved with respect to a real parameter,
 $\tau$. This implies that all quantities describing
the walls and their vacua indeed evolve with respect to $\tau$.
Then, the analysis on the eigenvalues of the first order expansion
of BPS equations shows that in general the vacua related to the
field theory on a curved background do not always exist. In order
to verify their existence in the ultraviolet or infrared regions
one has to perform the renormalization group analysis. Finally, we
discuss in detail a simple model with a linear superpotential and
the K\"ahler-Ricci soliton considered as the Rosenau solution.

\end{titlepage}




\section{Introduction}
Attempts to generalize the study of AdS/CFT correspondence
\cite{AdSCFT} on curved spacetimes have been done, for example in
the context of curved domain walls of five dimensional $N=2$
supergravity \cite{CDL1, CL}. In those papers the authors have
constructed curved BPS domain walls and discussed their dual
description in terms of the renormalization group (RG) flow
described by a beta function. By putting the supersymmetric field
theory studied in \cite{DWc} on a curved four dimensional AdS
background they have also demonstrated in a simple model that this
holographic RG flow in the field theory on the curved spacetime
has indeed a description in terms of curved BPS domain walls.\\
\indent So far, in the four dimensional $N=1$ supergravity theory
we only have the flat domain wall cases studied in some
references, for example \cite{FlatDW, RevFlatDW, CDGKL, GZ, GZA}.
Therefore, the above results motivate us to apply the scenario to
the case of the curved BPS domain walls of $N=1$ supergravity in
four dimensions. Our interest here is to study curved BPS domain
walls together with their Lorentz invariant vacua in the context
of the dynamical system and the RG flow analysis which is a
generalization of the previous works \cite{GZ, GZA}. Particularly,
we want to see how the general pattern looks like in a simple
model, namely the $N=1$ supergravity coupled to a chiral multiplet
whose scalar manifold can be regarded as a solution of the
K\"ahler-Ricci flow equation \cite{cao, cao1} \footnote{This
K\"ahler-Ricci flow also appears as one loop approximation of the
beta function of $N=2$ supersymmetry in two and three dimensions,
see for example \cite{HI, MN}. In this case, the parameter $\tau$
is regarded as the energy scale of the theory. However, this is
not the case in higher dimension, particularly in four
dimensions.}. This geometric soliton generates a one-parameter
family of scalar manifolds, \textit{i.e.} K\"ahler manifolds,
whose the deformation parameter is $\tau \in \lR$. In particular,
this K\"ahler-Ricci soliton can be viewed as a volume deformation
of a K\"ahler geometry for finite $\tau$ \footnote{This can be
easily seen if the initial geometry is a
K\"ahler-Einstein manifold. See Appendix \ref{RevKRF} for details.}.\\
 \indent Thus, defining $N=1$ supersymmetry on the K\"ahler-Ricci
 flow means that we deform it with respect to $\tau$. As a direct
 implication of such
treatment, all couplings such as the shifting quantities, the
masses of the fields, and the scalar potential do evolve with
respect to $\tau$ since those quantities depend on this geometric
soliton. Such behavior is generally inherited to all solitonic
solutions such as the domain walls. So, the Lorentz invariant
vacua do also possess such property which can shortly
be mentioned as follows.\\
 \indent First, near the vacua the
spacetime is in general non-Einsteinian and then, becomes a space
of constant curvature (which is also non-Einsteinian) related to
the divergences of the RG flow. Second, the K\"ahler-Ricci soliton
indeed affects the nature of the vacua\textbf{,} mapping
nondegenerate vacua to other degenerate vacua and vice versa.
Moreover, in a model that admits a singular geometric evolution,
the vacuum structure may have a parity pair of vacua in the sense
that the vacua of the index $\lambda$ turns into the other vacua
of the index $2 - \lambda$ after hitting the singularity. This is
an example that
also occurs in general the pattern for flat domain walls \cite{GZ}.\\
\indent Finally, in order to have a consistent picture, the
eigenvalues  of the first order expansion of the BPS equation have
to be real. This also shows that the above vacua do not always
exist in general. By performing the RG flow analysis we can
further verify the existence of such vacua in the infrared or
ultraviolet regions, correspond to the field theory on three
dimensional
curved background.\\
\indent The structure of this paper is as follows. In Section
\ref{CSKRS} we review the $N=1$ supergravity on a two dimensional
K\"ahler-Ricci soliton and introduce some quantities which are
useful for our analysis. Then, the discussion is continued in
Section \ref{BPSDW} by addressing some aspects of the curved BPS
domain walls on the two dimensional K\"ahler-Ricci solitons.
Section \ref{DSV} is assigned to the discussion of the nature of
the supersymmetric Lorentz invariant vacua together with their
deformation on the K\"ahler-Ricci soliton. We put the discussion
about the deformation of the supersymmetric flow on curved
spacetime in Section \ref{GRG}. A simple example is then given in
Section \ref{MRS}. Finally, we summarize our results in Section
\ref{Conclu}.

\section{$4d \; N=1$ Chiral Supergravity on $2d$ K\"ahler-Ricci Soliton}
\label{CSKRS}
 In this section we provide a  review of the four
dimensional $N=1$ supergravity coupled to a chiral multiplet in
which the non-linear $\sigma$-model satisfies the K\"ahler-Ricci
flow equation defined below, which in turn implies that our $N=1$
supergravity is defined on a one-parameter family of K\"ahler
manifolds
deformed with respect to the real parameter $\tau$ \cite{GZ}.\\
 \indent The ingredients of the $N=1$
theory are a gravitational multiplet and a chiral multiplet. The
gravitational multiplet is composed of a vierbein $e^a_{\nu}$ and
a vector spinor $\psi_{\nu}$ where $a=0,...,3$ and $\nu=0,...,3$
are the flat and the curved indices, respectively. The member of
the chiral multiplet is  a complex scalar $z$
and a spin-$\frac{1}{2}$ fermion $\chi$.\\
\indent We then construct a general $N=1$ chiral supergravity
Lagrangian together with its supersymmetry transformation. This
construction can be found, for example, in \cite{susyor}
\footnote{For an excellent review of $N=1$ supergravity, see also
for example \cite{susy, DF}}. Let us assemble the terms which are
useful for our analysis. The bosonic part of the $N=1$
supergravity Lagrangian has the form
\begin{equation}
{\mathcal{L}}^{N=1} = -\frac{M^2_P}{2}R + g_{z\bar{z}}(z, \bar{z};
\tau)\,
\partial_{\nu} z \,
\partial^{\nu}\bar{z} - V(z,\bar{z}; \tau)\;, \label{L}
\end{equation}
where $M_P$ is the Planck mass and by setting  $M_P \to +\infty$,
the $N=1$ global supersymmetric theory can be obtained. Next, $R$
is the Ricci scalar of the four dimensional spacetime; the pair
$(z,\bar{z})$ spans a Hodge-K\"ahler manifold with metric
$g_{z\bar{z}}(z, \bar{z}; \tau) \equiv
\partial_{z}
\partial_{\bar{z}}K(z,\bar{z}; \tau)$; and $K(z,\bar{z}; \tau)$
is a real function, called the K\"ahler potential. The scalar
manifold satisfies the K\"ahler-Ricci flow equation
\footnote{Classification of the special solutions of (\ref{KRF2d})
using linearization method has been studied, for example, in
\cite{CV}.}
\begin{equation}
\frac{\partial g_{z\bar{z}}}{\partial \tau} = -2
R_{z\bar{z}}(\tau) = -2 \;\partial_z
\bar{\partial}_{\bar{z}}{\mathrm{ln}}g_{z\bar{z}}(\tau)\;,
\label{KRF2d}
\end{equation}
where $\tau$ is a real parameter related to the deformation of a
K\"ahler surface mentioned in the previous section. The $N=1$
scalar potential $V(z,\bar{z}; \tau)$ has the form
\begin{equation}
 V(z,\bar{z}; \tau) = e^{K(\tau)/M^2_P}\left(g^{z\bar{z}}(\tau)\nabla_z W\,
 \bar{\nabla}_{\bar{z}} \bar{W}
 - \frac{3}{M^2_P} W \bar{W} \right)\;,
\label{V}
\end{equation}
where $W$ is a holomorphic superpotential and $\nabla_z W\equiv
(dW/dz)+ (K_z(\tau)/M^2_P) W$.  The Lagrangian (\ref{L}) is
invariant under the following supersymmetry transformations up to
three-fermion terms \footnote{The symbol $D_{\nu}$ here is
different with the one in reference \cite{GZA}. $D_{\nu}$ here is
defined as $D_{\nu} \equiv \partial_{\nu}
-\frac{1}{4}\gamma_{ab}\,\omega^{ab}_{\nu}$.}
\begin{eqnarray}
\delta\psi_{1\nu} &=& M_P \left(D_{\nu}\epsilon_1 +
\frac{\mathrm{i}}{2}e^{K(\tau)/2M^2_P}\,W \gamma_{\nu} \epsilon^1
+ \frac{\mathrm{i}}{2M_P} Q_{\nu}(\tau)\epsilon_1 \right)\;, \nonumber\\
\delta\chi^z &=& {\mathrm{i}}\partial_{\nu} z \, \gamma^{\nu}
\epsilon^1
+ N^z(\tau) \epsilon_1 \quad, \label{susytr}\\
\delta e^a_{\nu} &=&  - \frac{{\mathrm{i}}}{M_P} \, (
\bar{\psi}_{1\nu} \, \gamma^a \epsilon^1
+ \bar{\psi}^1_\nu \, \gamma^a \epsilon_1 )\;,\nonumber\\
\delta z &=& \bar{\chi}^z \epsilon_1 \;,\nonumber
\end{eqnarray}
where $N^z(\tau) \equiv
e^{K(\tau)/2M^2_P}\,g^{z\bar{z}}(\tau)\bar{\nabla}_{\bar{z}}\bar{W}$,
 $g^{z\bar{z}}(\tau) = (g_{z\bar{z}}(\tau))^{-1}$, and the $U(1)$ connection
 $Q_{\nu}(\tau) \equiv  - \left(  K_z(\tau) \,\partial_{\nu}z -
K_{\bar{z}}(\tau)\, \partial_{\nu}\bar{z}\right)$. Here, we have
also defined $\epsilon_1 \equiv \epsilon_1(x,\tau)$. The flow
equation (\ref{KRF2d}) implies that, for example, the scalar
potential (\ref{V}) and the shifting quantity $N^z$ do evolve with
 respect to $\tau$ as \cite{GZ}
\begin{eqnarray}
\frac{\partial N^z(\tau)}{\partial \tau} &=& 2 R^z_{\;z}(\tau)
N^z(\tau)+ \frac{K_{\tau}(\tau)}{2M_P^2} N^z(\tau)+
g^{z\bar{z}}(\tau) \frac{K_{\bar{z}\tau}(\tau)}{M_P^2}
e^{K(\tau)/2M^2_P}\, \bar{W}\;, \label{dV1t}\\
 \frac{\partial V(\tau)}{\partial
\tau} &=& \frac{\partial N^z(\tau)}{\partial \tau} N_z(\tau)+
\frac{\partial N_z(\tau)}{\partial \tau} N^z(\tau) -
\frac{3K_{\tau}(\tau)}{M_P^2} e^{K(\tau)/M^2_P}\, \vert W \vert^2
\;, \nonumber
\end{eqnarray}
where $R^z_{\;z} \equiv g^{z\bar{z}} R_{z\bar{z}}$. Note that as
has been studied in \cite{topping, caozhu} it is possible that
such geometric flow (\ref{KRF2d}) has a singular point at finite
$\tau$. For example, this situation can be directly observed in a
model with K\"ahler-Einstein manifold as initial geometry
\footnote{See also Appendix \ref{RevKRF}.}. In this paper we
particularly consider the Rosenau solution of (\ref{KRF2d}) which
also admits such property in section \ref{MRS}. This special
solution was firstly constructed in \cite{PR}.

\section{Curved BPS Domain Walls on $2d$ K\"ahler-Ricci Soliton}
\label{BPSDW}

This section is assigned for the discussion of curved domain walls
 admitting partial Lorentz invariance. In particular, we consider curved BPS domain walls that
maintain half of the supersymmetry of the parental theory. As a
consequence, the background should be a three
dimensional AdS spacetime.\\
\indent Let us now consider the ground states which partially
break
 the Lorentz invariance, \textit{i.e.} the domain walls. The
starting point is to take the ansatz metric of the four
dimensional spacetime as
\begin{equation}
ds^2 = a^2(u,\tau)\,
g_{\underline{\lambda}\,\underline{\nu}}\,dx^{\underline{\lambda}}\,
dx^{\underline{\nu}} \,- du^2 \;,\label{DWans}
\end{equation}
where $\underline{\lambda},\,\underline{\nu}=0,1,2$, $a(u, \tau)$
is the warped factor, and the parameter $\tau$ is related to the
dynamics of the K\"ahler metric governed by (\ref{KRF2d}). The
metric $g_{\underline{\lambda}\,\underline{\nu}}$ describes a
three dimensional AdS spacetime. Therefore, the corresponding
components of the Ricci tensor  of the metric (\ref{DWans}) are
given by
\begin{eqnarray}
R_{\underline{\lambda}\,\underline{\nu}} &=&  \left[ \left(
\frac{a'}{a} \right)' + 3 \left( \frac{a'}{a} \right)^2
  - \frac{\Lambda_3 }{a^2}\right] a^2
  g_{\underline{\lambda}\,\underline{\nu}}\;,\nonumber\\
  R_{33} &=& -3 \left[ \left( \frac{a'}{a} \right)'
  +  \left( \frac{a'}{a} \right)^2 \right] \;, \label{Riccians}
\end{eqnarray}
and the Ricci scalar has the form
\begin{equation}
 R = 6 \left[ \left( \frac{a'}{a} \right)' + 2 \left( \frac{a'}{a} \right)^2
 \right] - \frac{3 \Lambda_3 }{a^2}\;,\label{Rans}
\end{equation}
where $a' \equiv \partial a/\partial u$ and $\Lambda_3$ is the
negative three dimensional cosmological constant. Writing the
supersymmetry transformation (\ref{susytr}) and setting
$\psi_{1\nu} =\chi^z =0$ on the background (\ref{DWans}), leads to
\begin{eqnarray}
\frac{1}{M_P} \delta\psi_{1u} &=& D_u \,\epsilon_1 +
\frac{\mathrm{i}}{2} e^{K(\tau)/2M^2_P}\,
W\gamma_u \epsilon^1 + \frac{\mathrm{i}}{2M_P} Q_u (\tau)\epsilon_1 \;, \nonumber\\
\frac{1}{M_P} \delta\psi_{1\underline{\nu}} &=&
D_{\underline{\nu}}\,\epsilon_1 +
\frac{1}{2}\gamma_{\underline{\nu}}\left(-\frac{a'}{a}\gamma_3
\epsilon_1 + {\mathrm{i}}e^{K(\tau)/2M_P^2}\, W \epsilon^1
\right)+
\frac{\mathrm{i}}{2M_P} Q_{\underline{\nu}}(\tau)\epsilon_1\;, \label{susytr1}\\
\delta\chi^z &=& {\mathrm{i}}\partial_{\nu} z \, \gamma^{\nu}
\epsilon^1 + N^z(\tau) \epsilon_1 \;. \nonumber
\end{eqnarray}
 Supersymmetry further
demands that the right hand side of equations (\ref{susytr1})
vanish on the ground states. Then, on the three dimensional AdS
spacetime there exists a Killing spinor \cite{dH}
\begin{equation}
D_{\underline{\nu}}\,\epsilon_1 +\frac{\mathrm{i}}{2} \ell
\hat{\gamma}_{\underline{\nu}}\,\epsilon^1 =0 \;,
\end{equation}
where $\ell \equiv \sqrt{-\Lambda_3 / 2}$. Here,
$\hat{\gamma}_{\underline{\nu}}$ means the gamma  matrices in the
three dimensional AdS spacetime, and therefore
$\gamma_{\underline{\nu}} = a \hat{\gamma}_{\underline{\nu}}$.
Thus by taking $z = z(u, \tau)$ the first equation in
(\ref{susytr1}) shows that $\epsilon_1$ depends on the spacetime
coordinates, while the second equation gives a projection equation
\begin{equation}
\frac{a'}{a}\gamma_3 \epsilon_1  = {\mathrm{i}}
\left(e^{K(\tau)/2M^2_P}\, W(z) - \frac{\ell}{a} \right)\epsilon^1
\;,\label{projector}
\end{equation}
which leads to \footnote{This warped factor $a$ is related to the
 c-function in the holographic correspondence \cite{DWc, CDGKL}.}
\begin{equation}
\frac{a'}{a}  = \pm \,\left\vert e^{K(\tau)/2M^2_P}\,  W(z) -
\frac{\ell}{a} \right\vert \;. \label{warp}
\end{equation}
So, the warped factor $a$ is indeed $\tau$ dependent which is
consistent with our ansatz (\ref{DWans}).  Next, since $z = z(u,
\tau)$ the third equation in (\ref{susytr1}) becomes simply
\begin{eqnarray}
z' &=& \mp  \, 2e^{{\mathrm{i}}
\theta(\tau)}\,g^{z\bar{z}}(\tau)\, \bar{\partial}_{\bar{z}}
{\mathcal{W}}(\tau) \;,\nonumber\\
\bar{z}' &=& \mp \, 2 e^{-{\mathrm{i}}
\theta(\tau)}\,g^{z\bar{z}}(\tau)\,
\partial_z{\mathcal{W}}(\tau) \;, \label{gfe}
\end{eqnarray}
where we have introduced the phase function $\theta(z,\bar{z};u,
\tau)$ via
\begin{equation}
e^{{\mathrm{i}} \theta(\tau)} = \frac{\left(1 - \ell
e^{-K(\tau)/2M^2_P}\, (aW)^{-1} \right)}{\left\vert 1 - \ell
e^{-K(\tau)/2M^2_P}\, (aW)^{-1} \right\vert} \;, \label{fase}
\end{equation}
and the real function
\begin{equation}
{\mathcal{W}}(z,\bar{z}; \tau) \equiv e^{K(\tau)/2M^2_P}\, \vert
W(z) \vert \;.
\end{equation}
Note that at $\theta =0$ the flat domain wall case is regained,
which corresponds to $\ell =0$. Thus, we have the gradient flow
equations (\ref{gfe}), called the BPS equations in a curved
spacetime. Another supersymmetric flow related to the analysis is
the renormalization group (RG) flow given by the beta functions
\begin{eqnarray}
 \beta(\tau) &\equiv& a \frac{\partial z}{\partial a} =
 - \frac{2e^{{\mathrm{i}} \theta(\tau)}}
 {\left\vert e^{K(\tau)/2M^2_P}\,  W(z) - \ell / a \right\vert}\,
 g^{z\bar{z}}(\tau)\, \bar{\partial}_{\bar{z}}
{\mathcal{W}}(\tau) \;,\nonumber\\
\bar{\beta}(\tau) &\equiv& a \frac{\partial \bar{z}}{\partial a} =
- \frac{2e^{-{\mathrm{i}} \theta(\tau)}}{\left\vert
e^{K(\tau)/2M^2_P}\, W(z) - \ell / a
\right\vert}\,g^{z\bar{z}}(\tau)\,
\partial_z{\mathcal{W}}(\tau) \;, \label{beta}
\end{eqnarray}
after using (\ref{warp}) and (\ref{gfe}). These functions give a
description of a conformal field theory (CFT) on the three
dimensional AdS spacetime. Therefore, the scalars behave as
coupling constants and the warped factor $a$ can be viewed as an
energy scale \cite{DWc, CDGKL, CDKV},
 and the scalar potential (\ref{V}) can be written down as
\begin{equation}
 V(z,\bar{z}; \tau) = 4\, g^{z\bar{z}}(\tau)\, \partial_z {\mathcal{W}}(\tau)\,
 \bar{\partial}_{\bar{z}} {\mathcal{W}}(\tau)
 - \frac{3}{M^2_P}\, {\mathcal{W}}^2(\tau) \;.
\label{V1}
\end{equation}
\indent To see the deformation of the domain walls clearly, we
have to consider a case where, the K\"ahler-Ricci flow has a
singularity at finite $\tau = \tau_0 < \infty$. Such a property
may cause a topological change of the scalar manifold. The
simplest example of this case is when the initial manifold is
K\"ahler-Einstein manifold which has been studied in \cite{GZ}.
Here, our interest is only to look at another non trivial solution
of (\ref{KRF2d}) where
\begin{equation}
g_{z\bar{z}}(\tau) =\left\{ \begin{array}{ll}
g_1(z, \bar{z}; \tau) &   ; \;\;\;   \tau < \tau_0 \;, \\
0  &    ;\;\;\; \tau = \tau_0 \;, \\
- g_2(z, \bar{z}; \tau) &  ; \;\;\;  \tau > \tau_0 \;,\\
\end{array}
 \right.\label{KRsolumum}
 \end{equation}
with $g_1(z, \bar{z}; \tau)$ and $g_2(z, \bar{z}; \tau)$ are both
positive definite functions. For the case at hand the $N=1$ theory
(\ref{L}) and the walls described by (\ref{gfe}) and (\ref{beta})
diverge at $\tau = \tau_0$. This singularity disconnects two
different theories, namely  for $\tau < \tau_0$ we have $N=1$
theory on positive definite metric, while for $\tau > \tau_0$ it
becomes $N=1$ theory on negative definite metric.  As we will see
in section \ref{MRS} the Rosenau solution also has
similar property and produces a singularity at $\tau_0 = 0$.\\
 \indent Concluding this section, we will look at the gradient flow equation (\ref{gfe}).
 The critical points of the equation (\ref{gfe}) satisfy the following
 conditions
\begin{equation}
\partial_z{\mathcal{W}} = \bar{\partial}_{\bar{z}}{\mathcal{W}}
=0 \quad, \label{critpo}
\end{equation}
which imply that the first derivative of the scalar potential
(\ref{V1}) vanishes. In other words, there is a correspondence
between the critical points of ${\mathcal{W}}(\tau)$ and the vacua
of the $N=1$ scalar potential $V(\tau)$ \footnote{The vacuum of
the scalar potential (\ref{V}) means a Lorentz invariant vacuum
(ground state). In the following we will mention Lorentz invariant
vacuum just as vacuum or ground state.}. Moreover, the
K\"ahler-Ricci flow, specifically the metric (\ref{KRsolumum}),
causes an evolution of these critical points (vacua) which are
characterized by the beta functions (\ref{beta}) splitting into
two region. If $a \to +\infty$
 we have an ultraviolet (UV) region, while if $a
\to 0$ we have an infrared (IR) region . Several aspects of the
vacuum and the supersymmetric flow deformation will be discussed
in detail in section \ref{DSV} and section \ref{GRG} respectively.

\section{Properties of the Supersymmetric Vacua}
\label{DSV}

Our attention in this section will be mainly drawn to the
discussion of the supersymmetric vacua of the theory described by
the scalar potential (\ref{V1}). As mentioned in the previous
section these vacua are related to the critical points of the real
function ${\mathcal{W}}(\tau)$ and changing with respect to
$\tau$. We firstly show that in general such critical points are
correspond to a four dimensional non-Einsteinian spacetime. Then,
a second order analysis of the vacua related to the critical
points of ${\mathcal{W}}(\tau)$ will be carried out. Note that the
discussion here is incomplete since it does not involve a
supersymmetric flow analysis which is provided in section
\ref{GRG}. A short review about critical points of surfaces is
given in Appendix \ref{Hess} which maybe useful for the analysis.\\
\indent Let us begin our discussion by mentioning that from
(\ref{critpo}) a critical point of the real function
${\mathcal{W}}(\tau)$, say $p_0$, is in general $p_0 \equiv
(z_0(\tau), \bar{z}_0(\tau))$ due to the geometric flow
(\ref{KRF2d}). Such a point exists in the asymptotic regions,
namely around $u \to \pm \infty$. The form of the scalar potential
(\ref{V1}) at $p_0$ is
\begin{equation}
 V(p_0; \tau) = - \frac{3}{M^2_P}\, {\mathcal{W}}^2(p_0; \tau)
 \equiv - \frac{3}{M^2_P}\, {\mathcal{W}}^2_0\;.
\label{V2}
\end{equation}
For the case of the flat domain walls discussed in \cite{GZA}
 the equation (\ref{V2}) can be viewed as the cosmological constant
  of the spacetime at the vacuum.
However, in general this is not the case. Therefore, we have to
 consider the behavior of the warped factor $a$
near the vacua, which is related to the shape of the spacetime.
Around $p_0$, the solution of equation (\ref{warp}) tends to
\footnote{In the limit of flat walls we have $\ell \to 0$ and $A_0
\to \pm \infty$.}
\begin{equation}
a(u, \tau)  = \frac{l}{{\mathcal{W}}^2_0} \pm \left(
\frac{\ell^2}{{\mathcal{W}}^2_0} -
\frac{l^2}{{\mathcal{W}}^4_0}\right)^{1/2} \left[A_0 \, e^{\pm
{\mathcal{W}}_0 u} - A^{-1}_0 e^{\mp {\mathcal{W}}_0 u}\right] \;,
\label{warprev}
\end{equation}
where $ l \equiv \ell \, e^{K(p_0; \tau)/2M^2_P}
\,{\mathrm{Re}}W(z_0)$ and $A_0 \ne 0$. Since $a$ is real, then
${\mathcal{W}}_0 > \vert l \vert / \ell$. Moreover, in this case
we have $\left(a'/a\right)' \ne 0$  near $p_0$. So, defining
\begin{equation}
\frac{a'}{a}  = \pm \,\left\vert e^{K(p_0; \tau)/2M^2_P}\,  W(z_0)
-\frac{\ell}{a} \right\vert \equiv \pm \,k  \;, \label{warp1}
\end{equation}
with $k \equiv k(u, \tau) \ge 0$, the Ricci tensor
(\ref{Riccians}) and the Ricci scalar (\ref{Rans}) become

\begin{eqnarray}
R_{\underline{\lambda}\,\underline{\nu}} &=&  \left( \pm k' + 3
k^2 + 2\ell^2 \, e^{\mp 2 \int k \, du} \right) e^{\pm 2 \int k \,
du}  g_{\underline{\lambda}\,\underline{\nu}}\;,\nonumber\\
  R_{33} &=& -3 \left(\pm k' + k^2 \right)  \;, \nonumber\\
R &=& \pm 6 k' + 12 k^2 + 6\ell^2 e^{\mp 2\int k \, du}  \;,
\label{Riccians1}
\end{eqnarray}
respectively, which confirms that in general the spacetime is
non-Einsteinian \footnote{It is important to notice that the
analysis using supersymmetric flows, namely the gradient and the
RG flows, shows that such a spacetime does not always correspond
to a CFT in three dimensional AdS, see the discussion in section
\ref{GRG}.}. Let us consider some special cases as follows. For
$\ell =0$ case, we have a four dimensional AdS spacetime for $k
\ne 0$ with $k'=0$ and the cosmological constant given by
(\ref{V2}), appeared in the flat domain walls. Next, it is
possible to have a case where $k=0$. Here, the spacetime is four
dimensional non-Einsteinian space of constant curvature, where
\begin{equation}
 e^{K(p_0; \tau)/2M^2_P}\, W(z_0) = \frac{\ell}{a} \;, \label{singcon}
\end{equation}
 or in other words,
\begin{equation}
  {\mathrm{Im}}W(z_0) =0 \;. \label{singcon1}
\end{equation}
 These facts tell us that the first order expansion
 of the beta function (\ref{beta}) would be ill
 defined \footnote{See equations (\ref{eigen1U}) and
 (\ref{eigen2U}) in the next section for a detail.}.
 Hence, this vacuum does not correspond to the CFT on
 a three dimensional AdS spacetime.\\
 \indent Another singularity could occur
 at $\tau = \tau_0$ which is caused by the divergence of the geometric flow
 (\ref{KRsolumum}). The detail of this aspect depends on the model in
 which both the form of the geometric flow and the superpotential
 are involved. Also, in this case some quantities would diverge. We give a
 simple model in section \ref{MRS}.\\
 \indent The last case is a static case
 in which $W(z_0) \ne 0$ and the $U(1)$ connection vanishes
 \footnote{Note that in the $W(z_0) = 0$ case the warped factor $a(u, \tau)$
 in (\ref{warprev}) becomes singular.}.
 This means that $p_0$ does not depend on $\tau$, but rather is
 determined by the holomorphic superpotential $W(z)$. In other words,
 $p_0$ is a critical point of $W(z)$. However,
 as we will see in the following, although $p_0$ static, the second order
 analysis does depend on $\tau$ because the geometric flow described by the K\"ahler
 potential $K(z,\bar{z}; \tau)$ is involved in the analysis. An example of this
 situation is discussed in section \ref{MRS}.\\
 \indent The second part of the discussion is to study the properties
of the critical points of the real function ${\mathcal{W}}(\tau)$
which is $\tau$ dependent in the second order analysis. Here, the
eigenvalues of its Hessian matrix are also $\tau$ dependent and
have the form \footnote{Note that since we have curved BPS domain
walls, this eigenvalue has a restriction coming from the
eigenvalues of the first order expansion of the gradient flow
(\ref{gfe}). Again, see section \ref{GRG} for a
detail.\label{footnote}}
\begin{equation}
 \lambda^{\mathcal{W}}_{1,2}(\tau) = \frac{g_{z\bar{z}}(p_0; \tau)}
 {M_P^2}{\mathcal{W}}_0
 \pm 2 \lvert \partial^2_z {\mathcal{W}}_0\rvert  \:,\label{eigenvalW}
 \end{equation}
 where
\begin{equation}
\partial^2_z {\mathcal{W}}_0 \equiv \frac{e^{K (p_0; \tau)/M^2_P}\,
\bar{W}(\bar{z}_0)}{2{\mathcal{W}}(p_0; \tau)} \left(
\frac{d^2W}{dz^2}(z_0) + \frac{K_{zz}(p_0; \tau)}{M_P^2}W(z_0)+
\frac{K_z(p_0; \tau)}{M_P^2}\frac{dW}{dz}(z_0)\right)\:.
\label{duaW}
\end{equation}
Since the metric $g_{z\bar{z}}(\tau)$ satisfies (\ref{KRsolumum}),
we split the discussion into two parts. First, in the interval
$\tau < \tau_0$  the metric is positive definite, namely
$g_{z\bar{z}}(p_0; \tau) = g_1(p_0; \tau)$, and the possible cases
for $p_0$ are a local minimum if
\begin{equation}
   \lvert \partial^2_z {\mathcal{W}}_0\rvert < \frac{1}{2M_P^2}
   g_1(p_0; \tau){\mathcal{W}}_0 \:,
\label{minW}
\end{equation}
 or a saddle if
\begin{equation}
   \lvert \partial^2_z {\mathcal{W}}_0\rvert > \frac{1}{2M_P^2}
   g_1(p_0; \tau){\mathcal{W}}_0
   \:.\label{saddW}
\end{equation}
Furthermore, $p_0$ turns out to be degenerate when
\begin{equation}
   \lvert \partial^2_z {\mathcal{W}}_0\rvert = \frac{1}{2M_P^2}
   g_1(p_0; \tau){\mathcal{W}}_0
   \:,\label{degW}
\end{equation}
holds. Second, for $\tau > \tau_0$ we have a negative definite
metric and $g_{z\bar{z}}(p_0; \tau) = -g_2(p_0; \tau)$. In this
region the conditions (\ref{minW}) and (\ref{saddW}) are modified
by replacing $g_1(p_0; \tau)$ with $g_2(p_0; \tau)$ which further
state that $p_0$ is either a local maximum for
\begin{equation}
   \lvert \partial^2_z {\mathcal{W}}_0 \rvert < \frac{1}{2M_P^2}
   g_2(p_0; \tau){\mathcal{W}}_0\:,
\label{maxW}
\end{equation}
 or a saddle for
\begin{equation}
   \lvert \partial^2_z {\mathcal{W}}_0\rvert > \frac{1}{2M_P^2}
   g_2(p_0; \tau){\mathcal{W}}_0
   \:.\label{sadd1W}
\end{equation}
 For degenerate case, we use the same procedure on (\ref{degW}).
 Since the point $p_0$ is dynamic with respect to $\tau$, we can then
summarize the above results as follows. A non degenerate critical
point can be changed into a degenerate critical point and vice
versa by the special geometric flow (\ref{KRsolumum}). Moreover,
this flow also affects the index of the critical points, namely
the critical points of index $\lambda$ turn to another critical
points of index $2 - \lambda$ after passing the singularity at
$\tau = \tau_0$. In other words, this is a parity transformation
of the Hessian matrix of ${\mathcal{W}}$ that maps a critical
point in $\tau < \tau_0$ to its parity partner in $\tau
> \tau_0$, such as a local minimum to
a local maximum and vice versa \cite{GZ}.\\
 \indent In the
following we finally perform general analysis on the
supersymmetric vacua of the scalar potential (\ref{V1}) and their
relation to the critical points of ${\mathcal{W}}(\tau)$. As has
been mentioned in the preceding section, the critical point $p_0$
of ${\mathcal{W}}(\tau)$ defines a vacuum of the theory. At $p_0$
the $\tau$ dependent eigenvalues of the Hessian matrix of the
scalar potential (\ref{V1}) are given by \footnote{We have similar
discussion as in footnote \ref{footnote}.}
\begin{eqnarray}
 \lambda^{V}_{1,2}(\tau) &=& -4 \left( \frac{g_{z\bar{z}}(p_0; \tau)}{M_P^4}
 {\mathcal{W}}_0^2 - 2 g^{z\bar{z}}(p_0; \tau) \lvert \partial^2_z
 {\mathcal{W}}_0 \rvert^2 \right)\nonumber\\
 && \pm \, 4 \frac{{\mathcal{W}}_0}{M_P^2} \lvert \partial^2_z
 {\mathcal{W}}_0\rvert    \:.\label{eigenvalV}
\end{eqnarray}
The first step is to look for $\tau < \tau_0$. Local minimum of
the scalar potential (\ref{V1}) exists if
\begin{equation}
   \lvert \partial^2_z {\mathcal{W}}_0\rvert > \frac{g_1(p_0; \tau)}{M_P^2}
   {\mathcal{W}}_0    \:.
   \label{minV}
\end{equation}
On the other side, local maximum is ensured by
\begin{equation}
   \lvert \partial^2_z {\mathcal{W}}_0\rvert < \frac{g_1(p_0; \tau)}{2M_P^2}
   {\mathcal{W}}_0
   \:,\label{maxV}
\end{equation}
while the inequality
\begin{equation}
  \frac{g_1(p_0; \tau)}{2M_P^2}{\mathcal{W}}_0  < \lvert \partial^2_z
  {\mathcal{W}}_0\rvert
  < \frac{g_1(p_0; \tau)}{M_P^2}{\mathcal{W}}_0
  \:.\label{saddV}
\end{equation}
shows the existence of a saddle point. The vacua become degenerate
if
\begin{eqnarray}
  \lvert \partial^2_z {\mathcal{W}}_0\rvert &=&
\frac{g_1(p_0; \tau)}{2M_P^2}{\mathcal{W}}_0 \;, \nonumber\\
\lvert \partial^2_z {\mathcal{W}}_0 \rvert &=& \frac{g_1(p_0;
\tau)}{M_P^2}{\mathcal{W}}_0
  \:.\label{degV}
\end{eqnarray}
Since the metric $g_{z\bar{z}}(p_0; \tau)$ is negative definite
for $\tau > \tau_0$, one gets a local maximum by replacing
$g_1(p_0; \tau)$ with $g_2(p_0; \tau)$ in (\ref{minV}). By using
the same procedure to (\ref{maxV}) and (\ref{saddV}), conditions
for a local minimum and a saddle point are obtained, respectively.
Again, we have degenerate vacua if
\begin{eqnarray}
  \lvert \partial^2_z {\mathcal{W}}_0\rvert &=&
\frac{g_2(p_0; \tau)}{2M_P^2}{\mathcal{W}}_0 \;, \nonumber\\
\lvert \partial^2_z {\mathcal{W}}_0\rvert &=& \frac{g_2(p_0;
\tau)}{M_P^2}{\mathcal{W}}_0
  \:.\label{deg1V}
\end{eqnarray}
\indent Here, some comments are in order. In $\tau < \tau_0$
region, the analysis of equations (\ref{minW})-(\ref{degW}) and
(\ref{minV})-(\ref{degV}) gives the same results as in \cite{GZA}.
The local minimum of ${\mathcal{W}}(\tau)$, given by (\ref{minW}),
is mapped into the local maximum of the scalar potential
(\ref{maxV}). The other vacua, namely the local minimum, the
saddle point, and the degenerate case, described by the second
equality in (\ref{degV}), are coming from the saddle of
${\mathcal{W}}(\tau)$. Lastly, there is a case where both
eigenvalues of the Hessian matrix of the scalar potential and the
real function ${\mathcal{W}}(\tau)$ become singular. This means
that this degenerate critical point of ${\mathcal{W}}(\tau)$ is
mapped into the degenerate vacuum and
such a case is called intrinsic degenerate vacuum.\\
\indent For $\tau > \tau_0$ case, since the parity transformation
exists, we have the following situations. The existence of the
local minimum of the scalar potential is guaranteed by the local
maximum of ${\mathcal{W}}(\tau)$ given by (\ref{maxW}). The saddle
of ${\mathcal{W}}(\tau)$ are mapped into the local maximum, the
saddle point, and the degenerate vacua given by the second
equation in (\ref{deg1V}) of the scalar potential. But still, the
first equation in (\ref{deg1V}) describes intrinsic degenerate
vacua.

\section{Supersymmetric Flows On a Curved Spacetime}
\label{GRG}
 This section provides the analysis of the supersymmetric flows,
 namely the gradient flow equations (\ref{gfe}) and the RG flow described
 by the beta function (\ref{beta}) around the vacuum in the presence of geometric
 soliton (\ref{KRsolumum}). As we will see, such a soliton affects
 the flows by a minus sign which  is, in other words, the parity map
  mentioned above.\\
\indent First of all, we employ the dynamical system analysis on
the gradient flows (\ref{gfe}). A vacuum $p_0$ is an equilibrium
point of (\ref{gfe}) if it is also a critical point of
${\mathcal{W}}(\tau)$. Around $p_0$ the first order expansion of
(\ref{gfe}) gives the eigenvalues
\begin{eqnarray}
 \Lambda_{1,2} &=& \mp  \frac{{\mathcal{W}}_0}{M_P^2}
 \, {\mathrm{cos}}\theta_0(u, \tau) \nonumber\\
  && - 2 g^{z\bar{z}}(p_0; \tau) \left[ \lvert \partial^2_z
  {\mathcal{W}}_0\rvert^2 - \frac{g^2_{z\bar{z}}(p_0; \tau)}{4 M_P^4}
  {\mathcal{W}}_0^2 \, {\mathrm{sin}}^2\theta_0(u,
  \tau)\right]^{1/2}
 \label{eigengfe} \:,
\end{eqnarray}
where $\theta_0(u, \tau) \equiv \theta(p_0; u, \tau)$  and the
function $\theta(z, \bar{z}; u,\tau)$ is defined in (\ref{fase}).
In general the eigenvalues (\ref{eigengfe}) are complex because
the second term in the square root could be negative. Therefore in
order to have a consistent model we simply set that they must have
a real value in which \footnote{If we take the limit $\ell \to 0$,
then ${\mathrm{sin}}\theta(p_0;
  \tau) \to 0$. So, we regain the flat domain wall case with $\lvert \partial^2_z
  {\mathcal{W}}(p_0; \tau)\rvert \ge 0$ \cite{GZA}.}
\begin{equation}
\lvert \partial^2_z
  {\mathcal{W}}_0\rvert \ge  \frac{\lvert g_{z\bar{z}}
  (p_0; \tau) \rvert}{2 M_P^2}
  {\mathcal{W}}_0 \lvert {\mathrm{sin}}\theta_0(u,
  \tau)\rvert \;, \label{eigengfereal}
\end{equation}
holds. This inequality gives a restriction of the critical points
of the function ${\mathcal{W}}(\tau)$ and the vacua of the theory
described by (\ref{eigenvalW}) and (\ref{eigenvalV}),
respectively. In other words, in order to have some vacua related
to a CFT on the curved spacetime, the condition
(\ref{eigengfereal}) must be fulfilled \footnote{See also the
discussion on the RG flow below.}. \\
\indent For $\tau < \tau_0$ and ${\mathrm{cos}}\theta_0(u, \tau)
\ne 0$ case we obtain that the stable nodes require
\begin{equation}
  \lvert \partial^2_z {\mathcal{W}}_0\rvert >
\frac{g_1(p_0; \tau)}{2M_P^2}{\mathcal{W}}_0 \:, \label{node}
\end{equation}
while saddles are ensured by the condition
\begin{equation}
\frac{g_1(p_0; \tau)}{2 M_P^2}
  {\mathcal{W}}_0 \lvert {\mathrm{sin}}\theta_0(u,
  \tau)\rvert \le \lvert \partial^2_z {\mathcal{W}}_0 \rvert <
\frac{g_1(p_0; \tau)}{2M_P^2}{\mathcal{W}}_0 \:. \label{saddle}
\end{equation}
Looking at (\ref{node}), (\ref{minV}), and (\ref{saddV}) we find
that the dynamic of the walls described by (\ref{gfe}) is stable,
flowing along the local minimum and the stable direction
 of the saddles  of the scalar potential (\ref{V1}).
 Along local maximum the walls become
unstable and the gradient flow is an unstable saddle. In other
words, the dynamic of the walls is on an unstable curve flowing
away from $p_0$. Moreover, in this linear analysis there is also a
possibility of having $p_0$ as a bifurcation point, namely one of
the eigenvalue in (\ref{eigengfe}) vanishes. Such a case occurs if
 the condition (\ref{degW}) holds and it takes place on the intrinsic
 degenerate vacua\footnote{To see the type of this
 fold bifurcation one has to check the higher order terms. At least
 one of these terms is non vanishing \cite{dynsis}.}. These conclusions
  are similar as in the flat domain wall case \cite{GZA}.
  For ${\mathrm{cos}}\theta_0(u, \tau) = 0$, only the stable nodes
  survive and hence we have stable walls.\\
 \indent After crossing the singularity at $\tau = \tau_0$,
 \textit{i.e.} the $\tau > \tau_0$ case, we obtain the following inequality
\begin{equation}
  \lvert \partial^2_z {\mathcal{W}}_0 \rvert >
\frac{g_2(p_0; \tau)}{2M_P^2}{\mathcal{W}}_0 \:, \label{node1}
\end{equation}
describing unstable nodes, whereas saddles need
\begin{equation}
 \frac{g_2(p_0; \tau)}{2 M_P^2}
  {\mathcal{W}}_0 \lvert {\mathrm{sin}}\theta_0(u, \tau)\rvert
  \le \lvert \partial^2_z {\mathcal{W}}_0 \rvert <
\frac{g_2(p_0; \tau)}{2M_P^2}{\mathcal{W}}_0\:, \label{saddle1}
\end{equation}
assuming ${\mathrm{cos}}\theta_0(u, \tau) \ne 0$. We have unstable
walls flowing along the local maximum, and the unstable direction
of the saddles of the scalar potential (\ref{V1}). On the other
hand, along local minima the walls become stable approaching $p_0$
on the stable curve of the saddle flow. Again, a similar situation
is obtained for a bifurcation point which requires
\begin{equation}
  \lvert \partial^2_z {\mathcal{W}}_0\rvert =
\frac{g_2(p_0; \tau)}{2M_P^2}{\mathcal{W}}_0\:. \label{deg1W}
\end{equation}
If ${\mathrm{cos}}\theta_0(u,\tau) = 0$, then we have only
unstable nodes which means that the model admits only unstable
walls.\\
 \indent Now let us perform an analysis on the RG flows
described by the beta function (\ref{beta}) for finding out the
nature of the vacuum $p_0$ in the UV and  IR regions. Our starting
point is to expand the beta function (\ref{beta}) around $p_0$. We
obtain the matrix
\begin{equation}
{\mathcal{U}} \equiv - \left(
\begin{array}{ccc}
  \partial \beta/\partial z (p_0) & & \partial \bar{\beta}/
  \partial z (p_0) \\
  & & \\
  \partial \beta/\partial \bar{z}(p_0) & &
   \partial \bar{\beta} /\partial \bar{z}(p_0) \\
\end{array}
\right)\:, \label{matrixU}
\end{equation}
whose eigenvalues are
\begin{eqnarray}
\lambda^{\mathcal{U}}_1 &=& k^{-1} \Bigg(
\frac{{\mathcal{W}}_0}{M_P^2} {\mathrm{cos}} \theta_0(u,
\tau)  \nonumber\\
&& + \, 2 g^{z\bar{z}}(p_0; \tau) \left[ \lvert
\partial^2_z
  {\mathcal{W}}_0 \rvert^2 - \frac{g^2_{z\bar{z}}(p_0; \tau)}{4 M_P^4}
  {\mathcal{W}}_0^2 \, {\mathrm{sin}}^2\theta_0(u, \tau)\right]^{1/2} \Bigg) \:,
  \label{eigen1U}\\
\lambda^{\mathcal{U}}_2 &=&  k^{-1} \Bigg(
\frac{{\mathcal{W}}_0}{M_P^2} {\mathrm{cos}} \theta_0(u,
\tau) \nonumber\\
&& - \, 2 g^{z\bar{z}}(p_0; \tau) \left[ \lvert
\partial^2_z
  {\mathcal{W}}_0\rvert^2 - \frac{g^2_{z\bar{z}}(p_0; \tau)}{4 M_P^4}
  {\mathcal{W}}_0^2 \, {\mathrm{sin}}^2\theta_0(u,
  \tau)\right]^{1/2} \Bigg) \:. \label{eigen2U}
\end{eqnarray}
Similar as in the gradient flow, since the condition
(\ref{eigengfereal}) is fulfilled, then both eigenvalues
(\ref{eigen1U}) and (\ref{eigen2U}) are real. In the UV region
where $a \to +\infty$, it demands that at least one of the above
eigenvalues have to be positive along which the RG flow can depart
from the vacuum. Let us first discuss the $\tau < \tau_0$ region.
In ${\mathrm{cos}}\theta_0(u, \tau) > 0$ case, all possibilities
of the vacua are allowed. But, for ${\mathrm{cos}}\theta_0(u,
\tau) \le 0$ only stable nodes are permitted and therefore saddle,
local minimum, and non-intrinsic stable degenerate vacua could
exist. In other words, in this situation the gradient flow is only
flowing along the stable direction of such vacua. In $\tau
> \tau_0$ region we have the same result for
${\mathrm{cos}}\theta_0(u, \tau) > 0$ case. However, for
 ${\mathrm{cos}}\theta_0(u, \tau) \le 0$ case only unstable nodes
 are allowed which tell us the existence of unstable vacua such
  as saddle, local maximum, and non-intrinsic unstable degenerate
  vacua. In addition, the gradient flow is taking unstable
  direction of the vacua and hence we have here unstable
  situation.\\
\indent On the other side in the IR region, where $a \to 0$,
requires at least a negative eigenvalue of (\ref{matrixU}) which
is the direction of the RG flow approaching the vacuum. In $\tau <
\tau_0$ and ${\mathrm{cos}}\theta_0(u, \tau) \ge 0$ we find that
only stable nodes survive and again, we have only stable situation
in which local maximum and intrinsic degenerate vacua are
forbidden here. Conversely, we have all possible vacua
 for ${\mathrm{cos}}\theta_0(u, \tau) < 0$. Second, in $\tau > \tau_0$
 and ${\mathrm{cos}}\theta_0(u, \tau) \ge 0$ it turns out that
 only unstable nodes exist, therefore the theory admits only
 the unstable vacua mentioned above. Lastly, everything is allowed
 for ${\mathrm{cos}}\theta_0(u, \tau) < 0$.

\section{A Model with the Rosenau Soliton}
\label{MRS}

In this section we give an example in which the geometric flow is
called the Rosenau soliton satisfying (\ref{KRsolumum}). This
soliton was firstly studied in the context of fluid dynamics
\cite{PR} and has  been considered also by geometrician as a toy
model in two dimensional complex surfaces. For a review see for
example
\cite{chow}.\\
\indent The Rosenau soliton has the form
\begin{equation}
 g_{z\bar{z}}(\tau)= - \frac{4 c^2}{b}
 \frac{{\mathrm{sinh}}(
 2b\tau)}{{\mathrm{cosh}} \lbrack 2c(z+\bar{z})\rbrack +
 {\mathrm{cosh}}(2b\tau)} \;,
\label{Rosol}
\end{equation}
whose corresponding $U(1)$ connection is given by
\begin{equation}
 Q(\tau)=  - \frac{{\mathrm{i}}c}{b}\;
 {\mathrm{ln}}\!\!\left( \frac{{\mathrm{cosh}}\lbrack v
 + 2b\tau \rbrack}{{\mathrm{cosh}} v} \right)
  ( dz -d\bar{z}) \;,
\label{QRosol}
\end{equation}
where $c \in \lR $, $b>0$, and $v \equiv c(z+\bar{z})
 - b\tau$ which diverges at $\tau
=0$ or $c=0$. It is easy to see that the metric (\ref{Rosol}) is
invariant under parity transformation, $c \leftrightarrow -c$. We
can further get its K\"ahler potential
\begin{eqnarray}
K(\tau) &=&  - \frac{2}{b} \int {\mathrm{ln}}\!\!\left(
\frac{{\mathrm{cosh}}\lbrack v
 + 2b\tau \rbrack}{{\mathrm{cosh}} v} \right) dv  \nonumber\\
 &=& 2b \tau v + 2 \, {\mathrm{Re}}\Big\lbrack {\mathrm{dilog}}\!
 \left(1 + {\mathrm{i}} e^{v + 2b\tau } \right) - {\mathrm{dilog}}\!
 \left(1 + {\mathrm{i}} e^v \right)\Big\rbrack \label{KpotRos}\;,
\end{eqnarray}
where  dilogarithm function have been introduced \cite{AS}
\begin{equation}
{\mathrm{dilog}}(x) \equiv \int_1^x \frac{{\mathrm{ln}}\,t}{1-t}dt
 = \sum_{k=1}^{+ \infty} \frac{x^k}{k^2}\;.
\end{equation}
Looking at (\ref{Rosol}), we find that
\begin{equation}
 g_1(\tau)= g_2(\tau) =  \frac{4 c^2}{b}
 \frac{\vert {\mathrm{sinh}}(
 2b\tau) \vert}{{\mathrm{cosh}} \lbrack 2c(z+\bar{z})\rbrack +
 {\mathrm{cosh}}(2b\tau)} \;,
\end{equation}
for all $\tau$ but not at singularity $\tau =0$ and $c \ne 0$.\\
 \indent Now let us first choose for simplicity the
holomorphic superpotential
\begin{equation}
W(z) = a_0 + a_1 z \;, \label{linearsup}
\end{equation}
with $a_0, a_1 \in \lR$. So to find a supersymmetric vacuum one
has to solve the condition (\ref{critpo}), which in this model
becomes
\begin{equation}
 a_1 - \frac{2c}{b \, M_P^2}\;
 {\mathrm{ln}}\!\!\left( \frac{{\mathrm{cosh}}\lbrack c(z_0 + \bar{z}_0)
 + b\tau \rbrack}{{\mathrm{cosh}}\lbrack c(z_0 + \bar{z}_0)
 - b\tau \rbrack}\right) (a_0 + a_1 z_0) = 0 \;. \label{eqcritpo}
\end{equation}
Nondegeneracy requires $W(z_0) \ne 0$. We then obtain the solution
of (\ref{eqcritpo}) as
\begin{eqnarray}
y_0 &=&  0  \;, \nonumber\\
{\mathrm{tanh}}(2c x_0)\, {\mathrm{coth}} \left( \frac{b a_1
M_P^2}{4c(a_0 + a_1 x_0)}\right)&=& {\mathrm{coth}} (b\tau)\;,
\label{soleqcritpo}
\end{eqnarray}
which follow that the imaginary part of $W(z_0)$ vanishes . Thus,
from (\ref{singcon1}) we find that this model admits only singular
vacua which do not related to the CFT in three dimensions. In
addition, it is easy to see that the origin belongs to this class
of vacuum for $a_1 =0$ and $a_0 \ne 0$ at which the
 $U(1)$ connection disappears.\\
 \indent The next case is to replace $a_1$ by ${\mathrm{i}}a'_1$
 with $a'_1 \in \lR$ in the superpotential (\ref{linearsup}).
 Then, we obtain
\begin{eqnarray}
y_0 &=&  \frac{a_0}{a'_1}  \;, \nonumber\\
{\mathrm{tanh}}(2c x_0)\, {\mathrm{coth}} \left( \frac{b M_P^2}{4c
x_0}\right)&=& {\mathrm{coth}} (b\tau)\;, \label{soleqcritpo1}
\end{eqnarray}
in which the superpotential evaluated at $p_0$ has the form
\begin{equation}
W(z_0) =  {\mathrm{i}} a'_1 x_0 \;. \label{linearsup1}
\end{equation}
Hence, the warped factor (\ref{warprev}) simplifies to
\begin{equation}
a(u, \tau)  = \pm \frac{\ell}{{\mathcal{W}}_0}  \left[A_0 \,
e^{\pm {\mathcal{W}}_0 u} - A^{-1}_0 e^{\mp {\mathcal{W}}_0
u}\right] \;, \label{warprev1}
\end{equation}
where
\begin{equation}
{\mathcal{W}}_0 =  e^{K(x_0; \tau)/2M_P^2} \, \vert a'_1 x_0
\vert\;.
\end{equation}
We close this section by providing the rest quantities related to
the analysis of the eigenvalues, namely
$\lambda^{\mathcal{W}}_{1,2}, \lambda^V_{1,2}, \Lambda_{1,2},$ and
$ \lambda^{\mathcal{U}}_{1,2}$. These are
\begin{eqnarray}
\vert \partial^2_z {\mathcal{W}}_0\vert &=&
\frac{{\mathcal{W}}_0}{2M_P^2} \left\vert g_{z\bar{z}}(p_0; \tau)
- \frac{1}{M_P^2} \vert K_z \vert^2(p_0; \tau)  \right\vert \;, \nonumber\\
 {\mathrm{cos}}\theta_0(u, \tau) &=& \left(1+ \ell^2(a {\mathcal{W}}_0)^{-2}
 \right)^{-1/2} \;, \\
{\mathrm{sin}}\theta_0(u, \tau) &=& \pm \, \ell(a
{\mathcal{W}}_0)^{-1} \left(1+ \ell^2(a {\mathcal{W}}_0)^{-2}
\right)^{-1/2}\;. \nonumber
\end{eqnarray}
Therefore, we have a model where there may be a possibility of
having vacua which correspond to the CFT. Moreover, in this model
the singularities are at $\tau=0$ and at $a'_1=0$.

\section{Conclusions}
\label{Conclu}

We have studied the nature of the $N=1$ supergravity curved BPS
domain walls and their vacuum structure. In particular, we have
considered a four dimensional $N=1$ supergravity coupled to a
chiral multiplet whose the scalar manifold can be viewed as the
K\"ahler-Ricci soliton satisfying the geometric flow equation
(\ref{KRF2d}). Some consequences are emerged as follows. First of
all, all couplings such as the scalar potential and the shifting
quantity are evolving with respect to $\tau$ which are given in
(\ref{dV1t}). Next, the warped factor, the BPS equations, and the
beta function describing the domain walls and the three
dimensional CFT,
respectively, also depend on $\tau$.\\
\indent The analysis on the vacua of the theory shows that in
general the spacetime is non-Einsteinian, deformed with respect to
$\tau$ whose Ricci tensor and Ricci scalar have the form provided
in (\ref{Riccians1}). This corresponds to the CFT on the three
dimensional AdS spacetime which is ensured by the beta function
(\ref{beta}). In this paper we just mentioned three special cases.
First, for $\ell =0$ and $k \ne 0$ with $k'=0$ we regain four
dimensional AdS spacetimes which  appear in the flat domain walls.
Second, there is a singularity at $k=0$ and $\ell \ne 0$ related
to the RG flow analysis in which the spacetime has a constant
curvature but non-Einsteinian. Note that the singularity caused by
the geometric flow is trivial since some quantities in this case
would become singular. Third, if $W(z_0) \ne 0$ and the $U(1)$
connection vanishes, then we have a static case. For this case,
the vacua are defined by the critical points of the superpotential $W(z)$.\\
\indent Since our ground states generally depend on $\tau$ and
additionally the theory does admit the existence of singular point
at $\tau =\tau_0$ described in (\ref{KRsolumum}), then we have to
split the region in order to analyze the function
${\mathcal{W}}(\tau)$ and the scalar potential $V(\tau)$. In $\tau
< \tau_0$ region  we reproduce the previous results in \cite{GZA}
as follows. Our analysis confirms that the deformation of the
critical points of ${\mathcal{W}}(\tau)$ can only be in the
following types, namely local minima, saddles, and degenerate
critical points. The local minima verify the existence of local
maximum vacua, whereas the saddles are mapped into local minimum,
saddle, or non-intrinsic degenerate vacua. There is possibly a
special situation where we have intrinsic degenerate vacua coming
from the degenerate critical points of ${\mathcal{W}}(\tau)$.
Hence, these also prove that the vacua certainly deform with
respect to $\tau$.\\
 \indent For $\tau > \tau_0$ case, similar as mentioned above,
 ${\mathcal{W}}(\tau)$ also admits the evolution of its critical points
 which are local maxima, saddles, and degenerate critical points.
 In this case, however, the local maxima are mapped into local
 minimum vacua, while the saddles imply the existence of local
 maximum, saddle, or non-intrinsic degenerate vacua. In addition,
 intrinsic degenerate vacua could possibly exist. So, these results show
 that the geometric flow (\ref{KRsolumum}) indeed changes the
 Hessian matrix of the real function ${\mathcal{W}}(\tau)$
  by a minus sign. Or in other words, the vacua of the index $\lambda$ change
  to the other vacua of the index $2 - \lambda$ caused by K\"ahler-Ricci
  flow. The ground states of index $2 - \lambda$ in $\tau > \tau_0$ region
  are called the parity pair of those with the index $\lambda$
  in $\tau < \tau_0$ region.\\
 \indent Furthermore, the analysis using the gradient and the RG flows
 show that the above vacua do not always exist. First, the
 first order expansion of the BPS equations yields the condition
 (\ref{eigengfereal}) that ensures the existence of the vacua related to
 the three dimensional CFT. Then, in the interval $\tau <
 \tau_0$ we have stable nodes flowing along the local minimum, and
 the stable direction of the saddle vacua,
  whereas unstable saddles flow
 along the local maximum vacua. On the other side, for $\tau >
 \tau_0$ case, the gradient flow turns into the unstable nodes flowing
 along the local maximum and the unstable direction of the saddle
 vacua, while along the local minimum the gradient flow is flowing on the stable
 curve of the saddle flow which is called the stable saddles. In both
 region there is a possibility of having a bifurcation point
 occurring near intrinsic degenerate vacua.\\
 \indent In order to check the existence of a
 ground state one has to carry out the analysis on the beta function.
Particularly in the UV region and ${\mathrm{cos}}\theta_0(u, \tau)
\le 0$, for $\tau < \tau_0$ we have only stable nodes which
further imply that the theory admits only stable vacua, whereas
for $\tau > \tau_0$ the gradient flow turns into unstable nodes.
On the other hand, in the IR region and ${\mathrm{cos}}\theta_0(u,
\tau) \ge 0$, for $\tau < \tau_0$ again only stable situation is
allowed, while for $\tau > \tau_0$ everything becomes unstable.\\
\indent Next, we have considered a simple model in which the
superpotential is linear and the K\"ahler-Ricci soliton has a
special form called the Rosenau solution. In the model where $a_0,
a_1 \in \lR$ we find that the model has only singular ground
states unrelated to the three dimensional CFT. However, if we set
at least $ a_1 \in \lC$, we may obtain the vacua related to the
CFT.

\vskip 1truecm

\hspace{-0.6 cm}{\Large \bf Acknowledgement}
\\
\vskip 0.15truecm \hspace{-0.6 cm} \noindent  We thank K. Yamamoto
for the early stage of this work. We also acknowledge H. Alatas
and A. N. Atmaja for useful discussions related to the topics of
this paper. In addition, we particularly thank T. Mohaupt and M.
Satriawan for careful reading and correcting English grammar. We
are grateful to the people at the theoretical astrophysics group
and the elementary particle physics group of Hiroshima University
for warmest hospitality where the early stage of this work was
done. This work is supported by Riset KK ITB 2008 under contract
No.: 035/K01.7/PL/2008 and ITB Alumni Association (HR IA-ITB)
research project 2008 under contract No. 1241a/K01.7/PL/2008.

\appendix
\section{Convention and Notation}
The purpose of this appendix is to assemble  our conventions in
this paper. The spacetime metric is taken to have the signature
$(+,-,-,-)$ while the Christoffel symbol is given by
$\Gamma^{\mu}_{\nu\rho}=\frac{1}{2}g^{\mu\sigma}
(\partial_{\nu}g_{\rho\sigma}+\partial_{\rho}g_{\nu\sigma}
-\partial_{\sigma}g_{\nu\rho})$ where $g_{\mu\nu}$ is the
spacetime metric. The Riemann curvature has the form $R^{\mu}_{\;
\nu\rho\lambda} =\partial_{\rho}\Gamma^{\mu}_{\;\nu\lambda
}-\partial_{\lambda}\Gamma^{\mu}_{\;\nu\rho} +
\Gamma^{\sigma}_{\;\nu\lambda}\Gamma^{\mu}_{\;\sigma\rho} -
\Gamma^{\sigma}_{\;\nu\rho}\Gamma^{\mu}_{\;\sigma\lambda}$ and the
Ricci tensor is defined to be $R_{\nu\lambda} =
R^{\mu}_{\; \nu\mu\lambda}$.\\

The following indices are given:\\

\begin{tabular}{r @{\hspace{2.5 cm}}  l }
$\underline{\nu},\underline{\lambda} = 0,1,2$, & label
three dimensional curved spacetime indices \\
\\
$\underline{a}, \underline{b} = 0,1,2$, & label
three dimensional flat spacetime indices \\
\\
$\mu, \nu = 0,...,3$, & label four dimensional curved spacetime
indices \\
\\
$a, b = 0,...,3$, & label four dimensional flat spacetime indices \\
\\
\end{tabular}

\section{Critical Point of A Function}
\label{Hess}
 The structure and the logic of this section are similar to those
in \cite{GZA} which are useful for our analysis in the paper.
Firstly, we consider any arbitrary (real) $C^\infty$-function
$f(z,\bar{z})$. A critical point $p_0 = (z_0,\bar{z}_0)$ of
$f(z,\bar{z})$ satisfies
\begin{equation}
\frac{\partial f}{\partial z}(p_0)=0 \:,\quad \frac{\partial
f}{\partial \bar{z}}(p_0)=0 \:.
\end{equation}
The point $p_0$ is said to be a non-degenerate critical point if
the Hessian matrix of $f(z,\bar{z})$
\begin{equation}
H_f  \equiv 2 \left(
\begin{array}{ccc}
  \frac{\partial^2 f}{\partial z \partial \bar{z}}(p_0) & &
  \frac{\partial^2 f}{\partial z^2}(p_0) \\
  & & \\
  \frac{\partial^2 f}{\partial \bar{z}^2}(p_0) & &
  \frac{\partial^2 f}{\partial \bar{z} \partial z }(p_0) \\
\end{array}
\right)
\end{equation}
is non-singular, \textit{i.e.},
\begin{equation}
 {\textrm{det}}\,H_f = 4 \left[ \left(
\frac{\partial^2 f}{\partial z \partial \bar{z}}(p_0)\right)^2
-\frac{\partial^2 f}{\partial z^2}(p_0)\frac{\partial^2
f}{\partial \bar{z}^2}(p_0) \right] \ne 0 \: .
\end{equation}
The eigenvalues of the Hessian matrix $H_f$ are given by
\begin{eqnarray}
\lambda^f_1 &=& \frac{1}{2} \left( {\textrm{tr}}H_f +
\sqrt{\left({\textrm{tr}}H_f \right)^2-4\, {\textrm{det}}H_f
}\right) \nonumber \:,\\
\lambda^f_2 &=& \frac{1}{2} \left( {\textrm{tr}}H_f -
\sqrt{\left({\textrm{tr}}H_f \right)^2-4\, {\textrm{det}}H_f
}\right) \:.\label{eigenval}
\end{eqnarray}
\indent The eigenvalues defined in (\ref{eigenval}) can be
used to classify the critical point $p_0$ of the function $f$ as
follows:
\begin{itemize}
\item[1.] If $\lambda^f_1 > 0$ and $\lambda^f_2 > 0$, then $p_0$
is a local minimum describing a stable situation.
 \item[2.] If $\lambda^f_1 < 0$ and
$\lambda^f_2 < 0$, then $p_0$ is a local maximum describing an
unstable situation.
\item[3.] If $\lambda^f_1 > 0$ and
$\lambda^f_2 < 0$ or vice versa, then $p_0$ is a saddle point.
 \item[4.] If at least one of its eigenvalue vanishes, then $p_0$
is said to be degenerate.
\end{itemize}

\section{A Quick Review of the $2d$ K\"ahler-Ricci Flow}
\label{RevKRF}

This section is devoted to give a short review of the two
dimensional K\"ahler-Ricci flow equation
\begin{equation}
\frac{\partial g_{z\bar{z}}}{\partial \tau} = -2
R_{z\bar{z}}(\tau) = -2 \;\partial_z
\bar{\partial}_{\bar{z}}{\mathrm{ln}}g_{z\bar{z}}(\tau)\;,
\label{KRF2d1}
\end{equation}
discussed in Section \ref{CSKRS} where $\tau \in \lR$. Here, we
particularly consider a simple case where the initial geometry at
$\tau =0$ is a K\"ahler-Einstein surface satisfying
\begin{equation}
R_{z\bar{z}}(x, 0) = \Lambda_2 \, g_{z\bar{z}}(x;0) \;,
\label{KEM}
\end{equation}
where $g_{z\bar{z}}(x;0)$ is an Einstein metric and $\Lambda_2 \in
\lR$. Then, we show that a K\"ahler-Ricci soliton can be viewed as
an area deformation of a K\"ahler geometry for finite $\tau$.\\
 \indent Let us simply choose that the constant $\Lambda_2 >
0$ and the initial metric $g_{z\bar{z}}(x;0)$ is definite
positive. Taking the metric ansatz
\begin{equation}
g_{z\bar{z}}(x;\tau) = \rho(\tau) \, g_{z\bar{z}}(x;0) \;,
\label{ansatzmet}
\end{equation}
and then the definition of Ricci tensor, we have
\begin{equation}
R_{z\bar{z}}(x;\tau) = R_{z\bar{z}}(x;0) =  \Lambda_2 \,
g_{z\bar{z}}(x;0)\;.
\end{equation}
Inserting (\ref{ansatzmet}) into (\ref{KRF2d1}), we get
\begin{equation}
g_{z\bar{z}}(x;\tau) = (1-2\Lambda_2 \tau) \, g_{z\bar{z}}(x;0)
\;. \label{KRFsolKE}
\end{equation}
This geometric soliton has a singularity at $\tau = 1/2\Lambda_2$.
After hitting the singularity, namely for $\tau > 1/2\Lambda_2$,
the geometry changes such that the metric is negative definite
with a negative "cosmological" constant. This shows that the
K\"ahler-Ricci flow interpolates two different $N=1$ theories
disconnected by the singularity at $\tau = 1/2\Lambda_2$.\\
\indent Finally, we show that the K\"ahler-Ricci soliton can be
viewed as a volume deformation of a K\"ahler manifold for finite
$\tau$. To be precise, the area of the soliton (\ref{KRFsolKE})
has the form
\begin{equation}
\sqrt{{\mathrm{det}}g(x;\tau)} \, d^2 x = \vert 1-2\Lambda_2 \tau
\vert \sqrt{{\mathrm{det}}g(x;0)} \, d^2 x \;.
\end{equation}
For $0 \le \tau < 1/2\Lambda_2$, the geometry is diffeomorphic to
the initial geometry endowed with a positive definite metric with
$\Lambda_2 > 0$, while for $\tau > 1/2\Lambda_2$ one has a
geometry admitting a negative definite metric with negative
"cosmological" constant.

\end{document}